\documentclass[12pt]{article}
\usepackage{amsmath}
\usepackage{amssymb}
\usepackage{amsfonts}

\numberwithin{equation}{section}
\newcommand{\field}[1]{\mathbb{#1}}
\newcommand{\R}{\field{R}}
\newcommand{\bp}{b^{\dagger}}
\newcommand{\bb}{\bar{b}}
\newcommand{\bbp}{\bar{b}^{\dagger}}
\newcommand{\hv}{\hat{V}}
\newcommand{\hw}{\hat{W}}
\newcommand{\hy}{\hat{Y}}
\newcommand{\hh}{\hat{H}}
\newcommand{\hqq}{\hat{Q}}
\newcommand{\hq}{\hat{q}}
\newcommand{\hc}{\hat{C}}

\title{
Chiral super-Tremblay-Turbiner-Winternitz Hamiltonians and their dynamical superalgebra}
\author{C Quesne\\ 
{\small Physique Nucl\'eaire Th\'eorique et Physique Math\'ematique,  Universit\'e Libre de Bruxelles,} \\ 
{\small Campus de la Plaine CP229, Boulevard~du Triomphe, B-1050 Brussels, Belgium} \\
{\small E-mail: cquesne@ulb.ac.be}}
\date{ }
\begin{document}
\baselineskip=22pt plus 1pt minus 1pt
\maketitle

\begin{abstract} 
The family of Tremblay-Turbiner-Winternitz (TTW) Hamiltonians $H_k$ on a plane, corresponding to any positive real value of $k$, is shown to admit another ${\cal N} = 2$ supersymmetric extension than that previously introduced by the present author. This new extension is of the same kind as that considered by D'Hoker and Vinet in the study of magnetic monopoles and is characterized by the fact that all the irreducible representations of the corresponding ${\rm osp}(2/2, \R)$ dynamical superalgebra are atypical lowest-weight state ones. The new supersymmetric Hamiltonians may be referred to as chiral super-TTW Hamiltonians, the role of chirality being played here by the fermion number parity operator. 
\end{abstract}

\noindent
Keywords: quantum Hamiltonians, supersymmetric quantum mechanics, superalgebras, chirality

\noindent
PACS numbers: 03.65.Fd, 11.30.Pb
%
%
\newpage
\section{Introduction}

The recent discovery by Tremblay, Turbiner and Winternitz (TTW) \cite{tremblay09} of an infinite family of exactly solvable and integrable deformations of the harmonic oscillator on a plane
\begin{equation}
\begin{split}
  & H_k = - \partial_r^2 - \frac{1}{r} \partial_r - \frac{1}{r^2} \partial_{\varphi}^2 + \omega^2 r^2 +
       \frac{k^2}{r^2} [a(a-1) \sec^2 k\varphi + b(b-1) \csc^2 k\varphi], \\
  & 0 \le r < \infty, \qquad 0 \le \varphi < \frac{\pi}{2k},
\end{split}  
\end{equation}
has aroused a lot of interest due to their proof of the superintegrability of $H_k$ for $k=1$, 2, 3, 4 and their conjecture of an extension of this property to higher integer $k$ values. Their subsequent demonstration of the periodicity of bounded trajectories in the classical case for rational $k$ \cite{tremblay10} actually supported such a conjecture.\par
%
%
The classical superintegrability was then established for rational $k$ \cite{kalnins09} while the quantum one was demonstrated for odd $k$ \cite{cq10a}. A constructive proof of the superintegrability of the quantum system for rational $k$ was finally provided \cite{kalnins10a}.\par
%
%
On the other hand, the TTW system was shown to admit a three-dimensional classical generalization, which is superintegrable for rational $k$ \cite{kalnins10b}. Moreover it was converted via coupling constant metamorphosis into an infinite family of superintegrable deformations of the Coulomb potential on the plane with a simultaneous transformation of the integrals of motion \cite{post}.\par
%
%
In a different line of thought, two additional extensions were also proposed. The first one \cite{cq10b}, based on a dihedral group $D_{2k}$ and a Dunkl operator formalism, proved useful to demonstrate the superintegrability of $H_k$ for odd $k$ \cite{cq10a}. The second one \cite{cq10c}, employing two independent pairs of fermionic operators and an $osp(2/2, \R)$ superalgebraic formalism, provided a supersymmetric extension of $H_k$ of the same kind as the familiar Freedman and Mende super-Calogero model \cite{freedman} and its generalization to other Calogero-like models (see \cite{brink} and other references quoted in \cite{cq10c}). A connection between the dihedral and supersymmetric extensions was also established \cite{cq10d}.\par
%
%
The purpose of this paper is to show that the TTW system admits another supersymmetric extension based on a different $osp(2/2, \R)$ superalgebra from that introduced in \cite{cq10c} (henceforth to be referred to as I and whose equations will be quoted by their number preceded by I). Such a nonstandard approach was initiated by D'Hoker and Vinet in the study of a spinning particle in the presence of a magnetic monopole and a $\lambda^2/r^2$ potential \cite{dhoker}. Later on, it was shown by Ghosh \cite{ghosh} to be also relevant to many-particle supersymmetric quantum mechanics and to provide a so-called chiral super-Calogero model, associated with atypical irreducible representations (irreps) of an $osp(2/2, \R)$ superalgebra \cite{nahm, scheunert77a, scheunert77b, balantekin, frappat}. Hence this alternative supersymmetric extension is also termed chiral or atypical.\par
%
%
In section 2, we briefly review some results obtained in I for the TTW Hamiltonians standard supersymmetric extension. We then use them in section 3 to demonstrate the existence of another supersymmetric extension connected with some transformed $osp(2/2, \R)$ superalgebra. The irreps of the latter, as well as bases for them, are determined in section 4. Finally, section 5 contains the conclusion.\par
%
%
\section{\boldmath Standard $osp(2/2, \R)$ superalgebra for the TTW Hamiltonians}

Let us recall that the ${\rm osp}(2/2, \R)$ superalgebra is generated by eight operators, four even ones closing the ${\rm sp}(2, \R) \times {\rm so}(2)$ Lie algebra and four odd ones, which separate into two ${\rm sp}(2, \R)$ spinors \cite{nahm, scheunert77a, scheunert77b, balantekin, frappat}. In the Cartan-Weyl basis, the former can be written as $K_0$, $K_{\pm}$ and $Y$ while the latter are denoted by $V_{\pm}$ and $W_{\pm}$. They satisfy the following (nonvanishing) commutation or anticommutation relations
\begin{equation}
\begin{array}{ll}
  [K_0, K_{\pm}] = \pm K_{\pm}, &\qquad  [K_+, K_-] = - 2K_0, \\[0.3cm]
  [K_0, V_{\pm}] = \pm \tfrac{1}{2} V_{\pm}, &\qquad [K_0, W_{\pm}] = \pm \tfrac{1}{2} W_{\pm}, 
  \\[0.3cm]
  [K_{\pm}, V_{\mp}] = \mp V_{\pm}, &\qquad [K_{\pm}, W_{\mp}] = \mp W_{\pm}, \\[0.3cm]
  [Y, V_{\pm}] = \tfrac{1}{2} V_{\pm}, &\qquad [Y, W_{\pm}] = - \tfrac{1}{2} W_{\pm}, \\[0.3cm] 
  \{V_{\pm}, W_{\pm}\} = K_{\pm}, &\qquad \{V_{\pm}, W_{\mp}\} = K_0 \mp Y,      
\end{array}  \label{eq:com}
\end{equation}
together with the Hermiticity properties
\begin{equation}
  K_0^{\dagger} = K_0, \qquad K_{\pm}^{\dagger} = K_{\mp}, \qquad Y^{\dagger} = Y, \qquad 
  V_{\pm}^{\dagger} = W_{\mp}.  \label{eq:hermite}
\end{equation}
\par
%
%
As an important subsuperalgebra, $osp(2/2, \R)$ contains the $sl(1/1)$ superalgebra of standard supersymmetric quantum mechanics \cite{cooper}, whose generators may be defined by
\begin{equation}
  {\cal H}^s = 4 \omega (K_0 + Y), \qquad Q = 2 \sqrt{\omega}\, W_+, \qquad Q^{\dagger} = 2 \sqrt{\omega}\,
  V_-  \label{eq:susy}
\end{equation}
in terms of some convenient constant $\omega$ and satisfy the conventional (anti)commutation relations
\begin{equation}
  [{\cal H}^s, Q] = [{\cal H}^s, Q^{\dagger}] = 0, \qquad \{Q, Q^{\dagger}\} = {\cal H}^s.
\end{equation}
\par
%
%
In I, it was shown that the $osp(2/2, \R)$ generators can be realized in terms of the polar coordinates $r$, $\varphi$, the corresponding differential operators $\partial_r$, $\partial_{\varphi}$ and two pairs of fermionic creation and annihilation operators $(\bp_x, b_x)$ and $(\bp_y, b_y)$ or their `rotated' counterparts $(\bbp_x, \bb_x)$ and $(\bbp_y, \bb_y)$, defined by
\begin{equation}
  \bbp_x = \bp_x \cos \varphi + \bp_y \sin \varphi, \qquad \bbp_y = - \bp_x \sin \varphi + \bp_y \cos \varphi
\end{equation}
and similarly for $\bb_x$ and $\bb_y$. In such a realization, the $sp(2, \R)$ weight generator $K_0 = K_{0,{\rm B}} + \Gamma$, entering the definition of ${\cal H}^s$ in (\ref{eq:susy}), can be chosen in such a way that the purely bosonic part $4 \omega K_{0,{\rm B}}$ of this supersymmetric Hamiltonian reduces to the TTW Hamiltonian $H_k$. The $osp(2/2, \R)$ even generators then turn out to be given by
\begin{equation}
\begin{split}
  K_0 & = K_{0,{\rm B}} + \Gamma, \qquad K_{\pm} = K_{\pm,{\rm B}} - \Gamma, \\
  K_{0,{\rm B}} &= \frac{1}{4\omega} H_k, \qquad K_{\pm,{\rm B}} = \frac{1}{4\omega} [- H_k + 2 \omega^2
        r^2 \mp 2\omega (r\partial_r + 1)], \\ 
  \Gamma & = \frac{k}{2\omega r^2} \bigl\{a \bigl[\bbp_x \bb_x - \tan k\varphi \bigl(\bbp_x \bb_y 
       + \bbp_y \bb_x\bigr) + (k \sec^2 k\varphi - 1) \bbp_y \bb_y\bigr] \\
  & \quad + b \bigl[\bbp_x \bb_x + \cot k\varphi \bigl(\bbp_x \bb_y + \bbp_y \bb_x\bigr)
       + (k \csc^2 k\varphi - 1) \bbp_y \bb_y\bigr]\bigr\}, \\
  Y & = \frac{1}{2} \bigl[\bbp_x \bb_x + \bbp_y \bb_y - k(a+b) - 1\bigr]
\end{split}  \label{eq:even}
\end{equation}
while the corresponding odd generators can be expressed as
\begin{equation}
\begin{split}
  V_{\pm} & = \frac{1}{2\sqrt{\omega}} \biggl[\bbp_x \biggl(\mp \partial_r + \omega r \pm \frac{k(a+b)}{r}
         \biggr) \mp \bbp_y \frac{1}{r} (\partial_{\varphi} + ka \tan k\varphi - kb \cot k\varphi)\biggr], \\
  W_{\pm} & = \frac{1}{2\sqrt{\omega}} \biggl[\bb_x \biggl(\mp \partial_r + \omega r \mp \frac{k(a+b)}{r}
         \biggr) \mp \bb_y \frac{1}{r} (\partial_{\varphi} - ka \tan k\varphi + kb \cot k\varphi)\biggr].
\end{split}  \label{eq:odd}
\end{equation}
As a result, the supersymmetric extension of $H_k$ can be exhibited as
\begin{equation}
  {\cal H}^s = H_{k,{\rm B}} + H_{k,{\rm F}}, \qquad H_{k,{\rm B}} = H_k, \qquad H_{k,{\rm F}} = 4\omega
  (\Gamma + Y)
\end{equation}
with $\Gamma$ and $Y$ as in (\ref{eq:even}).\par
%
%
After multiplication by  the fermionic vacuum state $|0\rangle$, the wavefunctions
\begin{equation}
  \Psi_{N,n}(r, \varphi) = {\cal N}_{N,n} Z^{(2n+a+b)}_N (z) \Phi^{(a,b)}_n (\varphi), \qquad z = \omega r^2,
\end{equation}
of $H_k$, corresponding to the energy eigenvalues $E_{N,n} = 2\omega [2N + (2n+a+b)k + 1]$, $N$, $n=0$, 1, 2,~\ldots, and whose detailed expression can be found in equation (I4.1), yield eigenstates of ${\cal H}^s$, as well as of the $osp(2/2, \R)$ weight generators $K_0$ and $Y$, with eigenvalues ${\cal E}_{N,n} = 4\omega (N + nk)$, $\tau + N$ and $q$, respectively. Here $\tau$ and $q$ are defined by
\begin{equation}
  \tau = \left(n + \frac{a+b}{2}\right)k + \frac{1}{2}, \qquad q = - \frac{1}{2}[(a+b)k + 1].
\end{equation}
All the states
\begin{equation}
  |\tau, \tau + N, q\rangle = \Psi_{N,n} |0\rangle  \label{eq:irrep-1}
\end{equation}
with a definite value of $n$ (or $\tau$) and $N=0$, 1, 2,~\ldots\ span a $sp(2, \R)$ lowest-weight state (LWS) irrep characterized by $\tau$. On starting from such states, $osp(2/2, \R)$ irreps were built by completing their bases with one- and two-fermion states obtained from
\begin{align}
  & |+, \tau + N + \tfrac{1}{2}, q + \tfrac{1}{2}\rangle = [N + (n+a+b)k + 1]^{-1/2} V_+ |\tau, \tau + N, q
         \rangle,  \label{eq:+state} \\
  & |-, \tau + N - \tfrac{1}{2}, q + \tfrac{1}{2}\rangle = (N + nk)^{-1/2} V_- |\tau, \tau + N, q\rangle,
         \label{eq:-state} \\
  & |\pm, \tau + N, q + 1\rangle = [n (n+a+b)k^2]^{-1/2} V_+ V_- |\tau, \tau + N, q\rangle,
         \label{eq:+/-state}
\end{align}
whose explicit expressions can be derived from (I4.3).\par
%
%
It was shown that for any nonvanishing $n$ value, there is an $osp(2/2, \R)$ irrep $(\tau, q)$ decomposing into four $sp(2, \R) \times so(2)$ irreps $(\tau) (q)$, $\left(\tau - \frac{1}{2}\right) \left(q + \frac{1}{2}\right)$, $\left(\tau + \frac{1}{2}\right) \left(q + \frac{1}{2}\right)$ and $(\tau) (q+1)$. The basis states of the first one are given by (\ref{eq:irrep-1}) and those of the last three can be written as
\begin{align}
  & |\tau - \tfrac{1}{2}, \tau + N - \tfrac{1}{2}, q + \tfrac{1}{2}\rangle = \alpha_N |-, \tau + N - \tfrac{1}{2}, 
        q + \tfrac{1}{2}\rangle + \beta_N |+, \tau + N - \tfrac{1}{2}, q + \tfrac{1}{2}\rangle,  \label{eq:irrep-2} \\
  & |\tau + \tfrac{1}{2}, \tau + N + \tfrac{1}{2}, q + \tfrac{1}{2}\rangle = \gamma_N |-, \tau + N + \tfrac{1}{2}, 
        q + \tfrac{1}{2}\rangle + \delta_N |+, \tau + N + \tfrac{1}{2}, q + \tfrac{1}{2}\rangle,  \label{eq:irrep-3} \\
  & |\tau, \tau + N, q + 1\rangle = |\pm, \tau + N, q + 1\rangle,  \label{eq:irrep-4} 
\end{align}
where the coefficients $\alpha_N$, $\beta_N$, $\gamma_N$ and $\delta_N$ can be found in I. As functions of $r$, $\varphi$ multiplied by some fermionic states, the basis states (\ref{eq:irrep-2}), (\ref{eq:irrep-3}) and (\ref{eq:irrep-4}) can be expressed as\footnote{It should be noted that equations (\ref{eq:irrep-2-bis}) and (\ref{eq:irrep-3-bis}) were not given in I and are a new result. They can be easily proved by combining the explicit forms of $\alpha_N$, $\beta_N$, $\gamma_N$, $\delta_N$, $|-, \tau + N - \tfrac{1}{2}, q + \tfrac{1}{2}\rangle$ and $|+, \tau + N + \tfrac{1}{2}, q + \tfrac{1}{2}\rangle$, provided in I, with some well-known relations satisfied by the Laguerre polynomials contained in $Z^{(2n+a+b)}_N (z)$ \cite{gradshteyn}.}
\begin{equation}
\begin{split}
  & |\tau - \tfrac{1}{2}, \tau + N - \tfrac{1}{2}, q + \tfrac{1}{2}\rangle = \left(\frac{N + (2n+a+b)k}{n (2n+a+b)}
        \right)^{1/2} \frac{{\cal N}_{N,n}}{\sqrt{\omega}} Z^{\left(2n+a+b-k^{-1}\right)}_N (z) \\
  & \quad \times \left[n \Phi^{(a,b)}_n(\varphi) \bbp_x + (n+a+b) \Phi^{(a+1,b+1)}_{n-1}(\varphi) \bbp_y
        \right] |0\rangle,
\end{split} \label{eq:irrep-2-bis}
\end{equation}
\begin{equation} 
\begin{split}
  & |\tau + \tfrac{1}{2}, \tau + N + \tfrac{1}{2}, q + \tfrac{1}{2}\rangle = - \left(\frac{n+a+b}{(2n+a+b)
        [N + (2n+a+b)k + 1]}\right)^{1/2} \sqrt{\omega} {\cal N}_{N,n} \\
  & \quad \times Z^{\left(2n+a+b+k^{-1}\right)}_N (z) \left[\Phi^{(a,b)}_n(\varphi) \bbp_x - 
        \Phi^{(a+1,b+1)}_{n-1}(\varphi) \bbp_y\right] |0\rangle,
\end{split} \label{eq:irrep-3-bis} 
\end{equation}
\begin{equation} 
  |\tau, \tau + N, q + 1\rangle = \left(\frac{n+a+b}{n}\right)^{1/2} {\cal N}_{N,n} Z^{(2n+a+b)}_N(z)
        \Phi^{(a+1,b+1)}_{n-1}(\varphi) \bbp_x \bbp_y |0\rangle.   
\end{equation}
The irrep $(\tau, q)$ is a typical one, associated with nonvanishing eigenvalues of the second- and third-order Casimir operators \cite{nahm, scheunert77a, scheunert77b, balantekin, frappat}. It has the peculiarity of not being a LWS irrep in contrast to many $osp(2/2, \R)$ irreps found in physical applications. The state $|\tau, \tau, q\rangle$, although annihilated  by $K_-$ and $W_-$, is indeed not destroyed by the superalgebra third lowering generator $V_-$.\par
%
%
{}For $n=0$, the situation looks rather different in the sense that the two one-fermion states (\ref{eq:+state}) and (\ref{eq:-state}) with the same eigenvalue $\tau + N + \frac{1}{2}$ of $K_0$ become equal while the two-fermion states (\ref{eq:+/-state}) vanish. One then gets an atypical irrep $(\tau, q)$ with $\tau = -q$, decomposing into only two $sp(2, \R) \times so(2)$ irreps $(\tau) (q)$ and $\left(\tau + \frac{1}{2}\right) \left(q + \frac{1}{2}\right)$ (with basis states $|\tau, \tau + N, q\rangle$ and $|\tau + \frac{1}{2}, \tau + N + \frac{1}{2}, q + \frac{1}{2}\rangle = |+, \tau + N + \frac{1}{2}, q + \frac{1}{2}\rangle$, respectively) and leading to vanishing eigenvalues of the Casimir operators. Furthermore, $(\tau, q)$ becomes a LWS irrep based on the ground state of ${\cal H}^s$.\par
%
%
In the next section, we will proceed to introduce an alternative $osp(2/2, \R)$ superalgebra for the TTW Hamiltonians, which will lead to some contrasting results for the corresponding irreps.\par
%
%
\section{\boldmath Nonstandard $osp(2/2, \R)$ superalgebra for the TTW Hamiltonians}

To go from the standard to the nonstandard superalgebra, we are going to leave the $sp(2, \R)$ generators $K_0$ and $K_{\pm}$ unchanged, but to modify the definition of the odd generators $V_{\pm}$, $W_{\pm}$ and of the purely fermionic even generator $Y$, which will now be denoted by $\hv_{\pm}$, $\hw_{\pm}$ and $\hy$, respectively. As a result, the supersymmetric extension of $H_k$ will now be
\begin{equation}
  \hat{{\cal H}}^s = H_{k,{\rm B}} + \hh_{k,{\rm F}}, \qquad H_{k,{\rm B}} = H_k, \qquad 
  \hh_{k,{\rm F}} = 4\omega (\Gamma + \hy),
\end{equation}
with some new supercharges $\hqq = 2 \sqrt{\omega}\, \hw_+$ and $\hqq^{\dagger} = 2 \sqrt{\omega}\, \hv_-$.\par
%
%
{}For such a purpose, it proves useful to introduce the fermion number parity operator $\Pi$, which may be defined as
\begin{equation}
  \Pi = (2\bp_x b_x - 1) (2\bp_y b_y - 1) = (2\bbp_x \bb_x - 1) (2\bbp_y \bb_y - 1)
\end{equation}
and plays the role of chirality $\gamma_5$ \cite{dhoker, ghosh} in the present problem. The projection operators on even- and odd-parity states are then given by
\begin{equation}
  \Pi^{\pm} = \tfrac{1}{2}(1 \pm \Pi).  \label{eq:projection}
\end{equation}
\par
%
%
{}For the new odd generators, let us assume the expressions
\begin{equation}
  \hv_{\pm} = \Pi^+ (V_{\pm} + W_{\pm}), \qquad \hw_{\pm} = \Pi^- (V_{\pm} + W_{\pm})  \label{eq:odd-bis}
\end{equation}
with $V_{\pm}$ and $W_{\pm}$ as in (\ref{eq:odd}). It can be easily checked that such operators satisfy the counterparts of relations (\ref{eq:com}) and (\ref{eq:hermite}) (or combinations of them) that do not involve the $so(2)$ generator. The modified version of the latter can then be determined so as to fulfil the last relation of (\ref{eq:com}), namely
\begin{equation}
  \hy = \tfrac{1}{2} [\{\hv_-, \hw_+\} - \{\hv_+, \hw_-\}].
\end{equation}
On inserting (\ref{eq:projection}) in (\ref{eq:odd-bis}) and using some properties of the old generators, we arrive at any one of the following equivalent forms
\begin{equation}
\begin{split}
  \hy &= \Pi (- V_+ V_- - W_+ W_- - V_+ W_- + V_- W_+ - Y) \\
  &= \Pi (- V_+ V_- - W_+ W_- + W_- V_+ - W_+ V_- + Y) \\
  &= \Pi (- V_+ V_- - W_+ W_- - V_+ W_- - W_+ V_- + K_0) \\
  &= \Pi (V_- V_+ + W_- W_+ + W_- V_+ + V_- W_+ - K_0), 
\end{split} \label{eq:even-bis}
\end{equation}
all of which will be employed in section 4. As a next step, it is straightforward to verify that the counterparts of the remaining relations in (\ref{eq:com}) and (\ref{eq:hermite}) are also fulfilled.\par
%
%
{}Finally the explicit expressions (\ref{eq:even}) and (\ref{eq:odd}) of the old generators can be merged in definitions (\ref{eq:odd-bis}) and (\ref{eq:even-bis}) of the new ones to yield for the latter
\begin{equation}
\begin{split}
  \hv_{\pm} & = \frac{1}{2\sqrt{\omega}} \Pi^+ \biggl[\bbp_x \biggl(\mp \partial_r + \omega r \pm 
         \frac{k(a+b)}{r} \biggr) \mp \bbp_y \frac{1}{r} (\partial_{\varphi} + ka \tan k\varphi - kb \cot k\varphi)
         \\
  & \quad + \bb_x \biggl(\mp \partial_r + \omega r \mp \frac{k(a+b)}{r}\biggr) \mp \bb_y \frac{1}{r}
         (\partial_{\varphi} - ka \tan k\varphi + kb \cot k\varphi)\biggr],
\end{split} \label{eq:odd-ter-1}
\end{equation}
\begin{equation}
\begin{split}
  \hw_{\pm} & = \frac{1}{2\sqrt{\omega}} \Pi^- \biggl[\bbp_x \biggl(\mp \partial_r + \omega r \pm 
         \frac{k(a+b)}{r} \biggr) \mp \bbp_y \frac{1}{r} (\partial_{\varphi} + ka \tan k\varphi - kb \cot k\varphi)
         \\
  & \quad + \bb_x \biggl(\mp \partial_r + \omega r \mp \frac{k(a+b)}{r}\biggr) \mp \bb_y \frac{1}{r}
         (\partial_{\varphi} - ka \tan k\varphi + kb \cot k\varphi)\biggr]
\end{split} \label{eq:odd-ter-2}
\end{equation}
and
\begin{equation}
\begin{split}
  \hy & = \tfrac{1}{2} [- \bbp_x \bbp_y (\partial_{\varphi} + ka \tan k\varphi - kb \cot k\varphi) + \bb_y
         \bb_x (\partial_{\varphi} - ka \tan k\varphi + kb \cot k\varphi) \\
  & \quad + \bbp_x \bb_y (\partial_{\varphi} - ka \tan k\varphi + kb \cot k\varphi) - \bbp_y \bb_x 
         (\partial_{\varphi} + ka \tan k\varphi - kb \cot k\varphi) \\
  & \quad - (2\bbp_y \bb_y - 1) k(a+b) + \Pi].  
\end{split} \label{eq:even-ter}
\end{equation}
\par
%
%
It is worth observing that the transformation from the standard to the nonstandard $osp(2/2, \R)$ superalgebra has made the structure of some generators more intricate. In particular, the $so(2)$ generator, which was purely fermionic, has become an operator wherein the fermionic degrees of freedom are entangled with the bosonic angular ones. This will have far-reaching consequences for the irreps to be reviewed in the next section.\par
%
%
\section{\boldmath Irreducible representations of the nonstandard $osp(2/2, \R)$ superalgebra}

To characterize the irreps of the alternative $osp(2/2, \R)$ superalgebra and to build some bases for them, the simplest starting point consists in applying the new generators $\hy$, $\hv_{\pm}$ and $\hw_{\pm}$ to the basis states of the standard superalgebra irreps\footnote{It should be noted that since the $sp(2, \R)$ generators have not changed, the old basis states remain characterized by the same $sp(2, \R)$ irrep quantum number when considering the nonstandard superalgebra.}. For such a purpose, one may either use the expressions (\ref{eq:odd-bis}) and (\ref{eq:even-bis}) of the former in terms of the old generators and the known action of these on the basis states (algebraic method) or employ their expressions (\ref{eq:odd-ter-1}), (\ref{eq:odd-ter-2}) and (\ref{eq:even-ter}) in terms of differential operators acting on functions of $r$ and $\varphi$ together with some properties of Laguerre and Jacobi polynomials \cite{gradshteyn} (differential operator method). As a check, we have performed both types of calculations, but we shall limit ourselves here to the algebraic method, which is somewhat simpler.\par
%
%
Let us start with the $n \ne 0$ case. Since the $so(2)$ generator $\hy$ commutes with the fermion number parity operator $\Pi$, we may deal separately with even- and odd-parity states.\par
%
%
{}For the former, we get a mixing of zero- and two-fermion states because the first and second lines of equation (\ref{eq:even-bis}) imply that
\begin{equation}
\begin{split}
  & \hy |\tau, \tau + N, q\rangle = - (V_+ V_- + Y) |\tau, \tau + N, q\rangle \\
  & \quad = \tfrac{1}{2} [(a+b)k + 1] |\tau, \tau + N, q\rangle - [n(n+a+b)k^2]^{1/2} |\tau, \tau + N, q + 1
        \rangle 
\end{split}
\end{equation}
and
\begin{equation}
\begin{split} 
  & \hy |\tau, \tau + N, q + 1\rangle = - (W_+ W_- - Y) |\tau, \tau + N, q + 1\rangle \\
  & \quad = - [n(n+a+b)k^2]^{1/2} |\tau, \tau + N, q\rangle + \tfrac{1}{2} [- (a+b)k + 1] |\tau, \tau + N, 
        q + 1\rangle, 
\end{split}
\end{equation}
respectively. Diagonalizing the resulting $2 \times 2$ matrix leads for $\hy$ to the eigenvalues
\begin{equation}
  \hq_{\pm} = \tfrac{1}{2} [1 \pm (2n+a+b)k] \qquad \text{or} \qquad \hq_+ = \tau, \quad \hq_- = 
  - \tau + 1,  
\end{equation}
with corresponding eigenstates
\begin{equation}
\begin{split}
  & |\tau, \tau + N, \hq_{\pm}) = A_{\pm} |\tau, \tau + N, q\rangle + B_{\pm} |\tau, \tau + N, q + 1\rangle, \\
  & A_+ = B_- = \left(\frac{n+a+b}{2n+a+b}\right)^{1/2}, \qquad A_- = - B_+ = \left(\frac{n}{2n+a+b}\right)
         ^{1/2}.  
\end{split} \label{eq:new-even} 
\end{equation}
Note that in (\ref{eq:new-even}) we use round brackets to distinguish the eigenvectors of $\hy$ from those of $Y$, denoted by angular ones.\par
%
%
Considering next the odd-parity states, i.e., the one-fermion states, we get from the third and fourth lines of equation (\ref{eq:even-bis})
\begin{equation}
\begin{split}
  & \hy V_- |\tau, \tau + N, q\rangle = (W_+ W_- + V_+ W_- - K_0) V_- |\tau, \tau + N, q\rangle \\
  & \quad = - \bigl\{N + \bigl[n + \tfrac{1}{2}(a+b)\bigr]k \bigr\} V_- |\tau, \tau + N, q\rangle \\
  & \qquad + \{N [N + (2n+a+b)k]\}^{1/2} V_+ |\tau, \tau + N - 1, q\rangle  
\end{split}
\end{equation}
and 
\begin{equation}
\begin{split}
  & \hy V_+ |\tau, \tau + N - 1, q\rangle = - (W_- W_+ + V_- W_+ - K_0) V_+ |\tau, \tau + N - 1, q\rangle \\
  & \quad = - \{N [N + (2n+a+b)k]\}^{1/2} V_- |\tau, \tau + N, q\rangle \\
  & \qquad + \bigl\{N + \bigl[n + \tfrac{1}{2}(a+b)\bigr]k \bigr\} V_+ |\tau, \tau + N - 1, q\rangle,  
\end{split}
\end{equation}
respectively. On using equations (\ref{eq:+state}), (\ref{eq:-state}), (\ref{eq:irrep-2}), (\ref{eq:irrep-3}) and the explicit expressions of $\alpha_N$, $\beta_N$, $\gamma_N$, $\delta_N$, it is then a simple matter to convert these apparently complicated results into a single relation
\begin{equation}
\begin{split}
  & \hy \big | \tau \pm \tfrac{1}{2}, \tau + N \pm \tfrac{1}{2}, q + \tfrac{1}{2} \big\rangle = 
         \pm \bigl[n + \tfrac{1}{2}(a+b)\bigr]k \big | \tau \pm \tfrac{1}{2}, \tau + N \pm \tfrac{1}{2}, q +
         \tfrac{1}{2} \big\rangle \\
  & \quad = \bigl(\hq_{\pm} - \tfrac{1}{2}\bigr) \big | \tau \pm \tfrac{1}{2}, \tau + N \pm \tfrac{1}{2}, q 
         + \tfrac{1}{2} \big\rangle,
\end{split}
\end{equation}
showing that the old one-fermion states remain eigenvectors of the new $so(2)$ generator. In the round bracket notation, we may therefore write
\begin{equation}
  \big | \tau \pm \tfrac{1}{2}, \tau + N \pm \tfrac{1}{2}, \hq_{\pm} - \tfrac{1}{2} \bigr) =
  \big | \tau \pm \tfrac{1}{2}, \tau + N \pm \tfrac{1}{2}, q + \tfrac{1}{2} \big\rangle.  \label{eq:new-odd}
\end{equation}
\par
%
%
To complete the calculations, it only remains to apply the new odd generators $\hv_{\pm}$, $\hw_{\pm}$ to the even-parity states $|\tau, \tau + N, \hq_{\pm})$, defined in (\ref{eq:new-even}), since their action on the odd-parity ones will then be determined  by Hermiticity. From the presence of $\Pi^+$ in definition (\ref{eq:odd-bis}) of $\hv_{\pm}$, it directly follows that
\begin{equation}
  \hv_{\pm} |\tau, \tau + N, \hq_+) = \hv_{\pm} |\tau, \tau + N, \hq_-) = 0. 
\end{equation}
On the other hand, as a consequence of (\ref{eq:odd-bis}) and (\ref{eq:new-even}), we get
\begin{equation}
  \hw_+ |\tau, \tau + N, \hq_+) = A_+ V_+ |\tau, \tau + N, q\rangle + B_+ W_+ 
  |\tau, \tau + N, q + 1\rangle.
\end{equation}
From (\ref{eq:+state}), it results that $V_+ |\tau, \tau + N, q\rangle$ is proportional to $\big |+, \tau + N + \frac{1}{2}, q + \frac{1}{2}\rangle$. Furthermore, $W_+ |\tau, \tau + N, q + 1\rangle$ can be written as a linear combination of $\big |+, \tau + N + \frac{1}{2}, q + \frac{1}{2}\bigr\rangle$ and $\big |-, \tau + N + \frac{1}{2}, q + \frac{1}{2}\bigr\rangle$ by successively using (\ref{eq:irrep-4}), (\ref{eq:+/-state}), (\ref{eq:com}), (\ref{eq:+state}) and (\ref{eq:-state}). Equations (\ref{eq:irrep-3}) and (\ref{eq:new-odd}) then yield
\begin{equation}
  \hw_+ |\tau, \tau + N, \hq_+) = - [N + (2n+a+b)k + 1]^{1/2} \big |\tau + \tfrac{1}{2}, \tau + N + \tfrac{1}{2},
  \hq_+ - \tfrac{1}{2}\bigr).
\end{equation}
Similarly, it is straightforward to prove that
\begin{align}
  & \hw_- |\tau, \tau + N, \hq_+) = - \sqrt{N}\, \big |\tau + \tfrac{1}{2}, \tau + N - \tfrac{1}{2}, \hq_+ 
         - \tfrac{1}{2}\bigr), \\
  & \hw_+ |\tau, \tau + N, \hq_-) = \sqrt{N+1}\, \big |\tau - \tfrac{1}{2}, \tau + N + \tfrac{1}{2}, \hq_- 
         - \tfrac{1}{2}\bigr), \\
  & \hw_- |\tau, \tau + N, \hq_-) = [N + (2n+a+b)k]^{1/2} \big |\tau - \tfrac{1}{2}, \tau + N - \tfrac{1}{2}, 
        \hq_- - \tfrac{1}{2}\bigr).  
\end{align}
\par
%
%
Hence, for every nonvanishing value of $n$, the nonstandard $osp(2/2, \R)$ superalgebra has two irreps corresponding to $\hq_+ = \tau$ and $\hq_- = - \tau + 1$, respectively. They decompose into only two $sp(2, \R) \times so(2)$ irreps: $(\tau) (\hq_+)$, $\bigl(\tau + \frac{1}{2}\bigr) \bigl(\hq_+ - \frac{1}{2}\bigr)$ for the former and $(\tau) (\hq_-)$, $\bigl(\tau - \frac{1}{2}\bigr) \bigl(\hq_- - \frac{1}{2}\bigr)$ for the latter\footnote{It is worth observing here that $\hw_+ \hw_- |\tau, \tau, \hq_{\pm}) = - \hw_- \hw_+ |\tau, \tau, \hq_{\pm}) = 0$ due to the presence of $\Pi^-$ in $\hw_{\pm}$.}. Furthermore, since
\begin{equation}
  \hw_- |\tau, \tau, \hq_+) = 0, \qquad \hv_- \big |\tau - \tfrac{1}{2}, \tau - \tfrac{1}{2}, \hq_- - \tfrac{1}{2}
  \bigr) = 0, 
\end{equation}
the states $|\tau, \tau, \hq_+)$ and $\big |\tau - \tfrac{1}{2}, \tau - \tfrac{1}{2}, \hq_- - \tfrac{1}{2}\bigr)$ are LWS of these two $osp(2/2, \R)$ irreps, which may therefore be characterized by $(\tau, \hq_+) = (\tau, \tau)$ and $\bigl(\tau - \frac{1}{2}, \hq_- - \frac{1}{2}\bigr) = \bigl(\tau - \frac{1}{2}, - \tau + \frac{1}{2}\bigr)$, respectively. The decomposition of $(\tau, \hq_+)$ and $\bigl(\tau - \frac{1}{2}, \hq_- - \frac{1}{2}\bigr)$ into subalgebra irreps entirely agrees with the general theory of atypical LWS irreps \cite{balantekin}.\par
%
%
Whenever $n$ vanishes, only the first atypical irrep survives and its basis states coincide with those of the atypical irrep previously found for the standard superalgebra, i.e.,
\begin{equation}
  |\tau, \tau + N, \hq) = |\tau, \tau + N, q\rangle, \qquad \big |\tau + \tfrac{1}{2}, \tau + N + \tfrac{1}{2}, \hq
  - \tfrac{1}{2}\bigr) = \big |\tau + \tfrac{1}{2}, \tau + N + \tfrac{1}{2}, q + \tfrac{1}{2}\bigr\rangle 
\end{equation}
with $\hq = \tau = - q$.\par
%
%
As predicted by the general theory, the second- and third-order Casimir operators of the nonstandard $osp(2/2, \R)$ superalgebra
\begin{equation}
\begin{split}
  \hc_2 & = (K_0 + \hy) (K_0 - \hy) - K_+ K_- - \hw_+ \hv_- - \hv_+ \hw_-, \\
  \hc_3 & = (K_0 + \hy) (K_0 - \hy) \hy - \bigl[\bigl(\hy + \tfrac{1}{2}\bigr) K_+ - \tfrac{1}{2} \hv_+ \hw_+
        \bigr] K_- + \tfrac{1}{2} (K_0 - 3\hy - 1) \hw_+ \hv_- \\
  & \quad + \tfrac{1}{2} [K_+ \hv_- - (K_0 + 3\hy - 1) \hv_+] \hw_-
\end{split}
\end{equation}
have vanishing eigenvalues in all its irreps.\par
%
%
\section{Conclusion}

In this paper, we have shown that by modifying the definition of some generators of the $osp(2/2, \R)$ dynamical superalgebra previously obtained for an ${\cal N} = 2$ supersymmetric extension ${\cal H}^s$ of the TTW Hamiltonians $H_k$, one can build another extension $\hat{\cal{H}}^s$ of the same with a corresponding transformed $osp(2/2, \R)$ dynamical superalgebra, which has only atypical LWS irreps. This new ${\cal N} = 2$ supersymmetric extension may therefore be termed atypical and $\hat{\cal{H}}^s$ may be referred to as a chiral super-TTW Hamiltonian, the role of chirality being played here by the fermion number parity operator $\Pi$.\par
%
%


\newpage
\begin{thebibliography}{99}

\bibitem{tremblay09} Tremblay F, Turbiner A V and Winternitz P 2009 {\em J.\ Phys.\ A: Math.\ Theor.} {\bf 42} 242001

\bibitem{tremblay10} Tremblay F, Turbiner A V and Winternitz P 2010 {\em J.\ Phys.\ A: Math.\ Theor.} {\bf 43} 015202

\bibitem{kalnins09} Kalnins E G, Miller W Jr and Pogosyan G S 2009 Superintegrability and higher order constants for classical and quantum systems arXiv:0912.2278 

\bibitem{cq10a} Quesne C 2010 {\em J.\ Phys.\ A: Math.\ Theor.} {\bf 43} 082001

\bibitem{kalnins10a} Kalnins E G, Kress J M and  Miller W Jr 2010 {\em J.\ Phys.\ A: Math.\ Theor.} {\bf 43} 265205 

\bibitem{kalnins10b} Kalnins E G, Kress J M and Miller W Jr 2010 {\em J.\ Phys.\ A: Math.\ Theor.} {\bf 43} 092001

\bibitem{post} Post S and Winternitz P 2010 {\em J.\ Phys.\ A: Math.\ Theor.} {\bf 43} 222001 

\bibitem{cq10b} Quesne C 2010 {\em Mod.\ Phys.\ Lett.} A {\bf 25} 15

\bibitem{cq10c} Quesne C 2010 {\em J.\ Phys.\ A: Math.\ Theor.} {\bf 43} 305202

\bibitem{freedman} Freedman D Z and Mende P F 1990 {\em Nucl.\ Phys.} B {\bf 344} 317

\bibitem{brink} Brink L, Turbiner A and Wyllard N 1998 {\em J.\ Math.\ Phys.} {\bf 39} 1285

\bibitem{cq10d} Quesne C 2010 {\em Mod.\ Phys.\ Lett.} A {\bf 25} 2373

\bibitem{dhoker} D'Hoker E and Vinet L 1985 {\em Commun.\ Math.\ Phys.} {\bf 97} 391

\bibitem{ghosh} Ghosh P K 2004 {\em Nucl.\ Phys.} B {\bf 681} 359

\bibitem{nahm} Nahm W and Scheunert M 1976 {\em J.\ Math.\ Phys.} {\bf 17} 868

\bibitem{scheunert77a} Scheunert M, Nahm W and Rittenberg V 1977 {\em J.\ Math.\ Phys.} {\bf 18} 146

\bibitem{scheunert77b} Scheunert M, Nahm W and Rittenberg V 1977 {\em J.\ Math.\ Phys.} {\bf 18} 155

\bibitem{balantekin} Balantekin A B, Schmitt H A and Halse P 1989 {\em J.\ Math.\ Phys.} {\bf 30} 274

\bibitem{frappat} Frappat L, Sciarrino A and Sorba P 1996 Dictionary on Lie superalgebras arXiv:hep-th/9607161

\bibitem{cooper} Cooper F, Khare A and Sukhatme U 1995 {\em Phys.\ Rep.} {\bf 251} 267

\bibitem{gradshteyn} Gradshteyn I S and Ryzhik I M 1980 {\em Table of Integrals, Series, and Products} (New York: Academic)

\end {thebibliography}
\end{document}